\documentclass[aps,prd,twocolumn,showpacs,superscriptaddress,groupedaddress]{revtex4}  
\usepackage{graphicx}  
\usepackage{dcolumn}   
\usepackage{bm}        
\usepackage{amsmath,amssymb}   

\usepackage{epstopdf}
\usepackage{hyperref}
\hypersetup{colorlinks=true, linkcolor=blue, urlcolor=blue, citecolor=red, bookmarks=false, pdfstartview={FitH}}

\begin{document}

\widetext

\title{Galaxy Rotation Curves via Conformal Factors}

\author{Ciprian A. Sporea}%
 \email{ciprian.sporea@e-uvt.ro}
\affiliation{\small \it
 West University of Timi\c soara, V.  P\^ arvan Ave.  4, RO-300223 Timi\c soara, Romania}
\author{Andrzej Borowiec}
 \email{andrzej.borowiec@ift.uni.wroc.pl}
 \author{Aneta Wojnar}%
 \email{aneta.wojnar@poczta.umcs.lublin.pl}
 \affiliation{\small \it
 Institute for Theoretical Physics, pl. M. Borna 9, 50-204, Wroclaw, Poland}

\begin{abstract}
Abstract:
  We propose a new formula to explain circular velocity profiles of spiral galaxies obtained from the Starobinsky model in Palatini formalism.
It is based on the assumption that the gravity can be
described by two conformally related metrics: one of them is responsible for the measurement of distances, while the other so-called dark
metric, is responsible for a geodesic equation and therefore can be used for the description of the velocity profile. The formula is tested
against a subset of galaxies taken from the HI Nearby Galaxy Survey (THINGS).
\end{abstract}

\maketitle

\section{Introduction and motivation}

Since there exist many issues that recently appeared in fundamental physics, astrophysics and cosmology which cannot be explained by General Relativity (GR) \cite{ein1, ein2}, one looks for other approaches which allow to understand their mechanism. Classical GR is a well-posed theory. Many astronomical observations tested GR and have confirmed that it is the best matching theory that we have had so far for explaining gravitational phenomena. Unfortunately, GR is not enough to describe many unsolved problems such as late-time cosmic acceleration \cite{hut, sami} (which one explains by existing an exotic fluid called Dark Energy introduced to the standard Einstein's field equations as cosmological constant), the Dark Matter puzzle \cite{kap, oort, zwi1, zwi2, bab, rub1, rub2, cap1, cap2}, inflation \cite{starob, guth}, as well as the renormalization problem \cite{stelle}.

There are two main ideas competing for an explanation of the Dark Matter problem: by the geometric modification of the gravitational field equations (see e.g. \cite{cap1, nodi, nodi2}) or by going beyond the Standard Model of Elementary Particles and introducing weakly interacting particles which have failed to be detected \cite{bertone}. In fact, these two ideas do not contradict each other and can be combined together in some future successful theory. The existence of Dark Matter is mainly indicated by anomalies in observed galactic rotation curves. It interacts only gravitationally with visible matter and radiation, and also has effects on the large-scale structure of the Universe \cite{davis, refre}.

Many interesting and promising models have faced the Dark Matter problem. The most famous one is Modified Newtonian Dynamics (MOND) \cite{millg, millg2, mond1,mond2,mond3,mond4,mc1, mc2}. It has predicted many galactic phenomena and hence it is widely used by astrophysicists. Closely related is the so-called Tensor/Vector/Scalar (TeVeS) theory of gravity \cite{bake, moffat1} which is, roughly speaking, the relativistic version of MOND. Another approach is to consider Extended Theories of Gravity (ETGs) - one modifies the geometric part of the field equations \cite{iorio, capzz, seba}. There are also attempts to obtain MOND result from ETGs, see for example \cite{barvinsky, fam, bernal, barientos, brun, fares}. Another interesting proposal for explaining rotation curves is by using the Weyl conformal gravity \cite{mannh1,mannh2,mannh3}. It should be noted that there  is a model based on a quantum effective action and large scale renormalization group effects \cite{rodr, rodr3}, later on constrained by the Solar System tests \cite{rodr2}.  It provides, up to the first nontrivial order, a conformal transformation of the spacetime metric and  a logarithmic term in the modified Newtonian
potential.

In the following paper we would like to show how the so-called "Dark Metric" $h_{\mu\nu}$  \cite{Fatibene, felicia}, that is, the metric which is conformally related to the physical metric $g_{\mu\nu}$ appearing in an action of a theory under consideration, may explain the galaxies rotation curves flatness problem. To this aim we employ the Ehlers-Pirani-Schild approach (EPS) \cite{EPS}
in the way it is pesented and considered  in Refs. \cite{Fatibene, felicia}. The formalism assumes that geometry of spacetime can be described by two structures, that is, conformal and projective ones. The first one is a class of Lorentzian metrics related to each other by the conformal transformation
\begin{equation}
 h=\Omega g,
\end{equation}
where $\Omega$ is a positive defined function (such transformation is often interpreted as a change of frame,  e.g. in scalar-tensor theories). The projective structure instead is a class of connections such that
\begin{equation}\label{connec}
 \tilde{\Gamma}^\alpha_{\beta\mu}=\Gamma^\alpha_{\beta\mu} +A_{(\mu}\delta^\alpha_{\beta)},
\end{equation}
with $A_\mu$ being a $1$-form. Because of positivity of the conformal function $\Omega$, the conformal structure defines light cones as well as timelike, lightlike and spacelike direction in considered spacetime. It should be noticed that it does not determine lengths of timelike and spacelike curves unless one chooses a representative of the conformal class. Geodesics in spacetime are defined by a connection. The different connections belonging to a considered projective structure define the same geodesics which are parameterized in two different ways \cite{Fatibene}. The choice of a parametrization is related to a choice of clock, that is, a metric.
We will say that the two discussed structures are EPS-compatible if the following holds (cf. (\ref{connec}))
\begin{equation}\label{condition}
 \tilde{\nabla}_\mu g_{\alpha\beta}=2A_\mu g_{\alpha\beta}.
\end{equation}
A triple consisting of the spacetime manifold $M$ and EPS-compatible structures is called EPS geometry \cite{Fatibene}.

We would like to emphasize that GR is a very special case of the described formalism. One assumes at the very beginning that the connection $\tilde{\Gamma}$ is a Levi-Civita connection of the metric $g$ (the $1$-form $A$ is zero) which results to treat the action of the theory as just metric-dependent. One may also treat the Einstein-Hilbert action as the one depending on two independent objects, that is, the metric $g$ and the connection $\tilde{\Gamma}$. This approach is called Palatini formalism. Considering the simplest gravitational Lagrangian, linear in scalar curvature $R$, Palatini approach leads to the dynamical result that $\tilde{\Gamma}$ is
a Levi-Civita connection of the metric $g$. It is not so in the case of more complicated Lagrangians appearing in ETGs. Moreover, as
it was shown in \cite{bernal2} that all Palatini connections of the form (\ref{connec}) are singled out by the variational principle.

Our aim is to show how important the EPS interpretation can be for an explanation of galaxy rotation curves. It turns out as expected that at the end we deal with the expression which consists of the Newtonian part and some modification which depends on the theory one wants to study under the EPS approach. As the simplest example, which we want to examine is  Starobinsky quadratic Lagrangian \cite{starob} in the Palatini formalism,    which currently  reaches very good results in the cosmological applications \cite{bor1, bor2, bor3}. The starting point will be the standard geodesic equation from which we will derive the rotational velocity. It will be shown that the velocity can be written as GR plus extra terms coming from the conformal factor.  Its usefulness is tested on  a sample of 6 HSB galaxies. The conclusions and future ideas will be drawn in the last part of the paper. The metric signature convention is $(-,+,+,+)$.

\section{Velocities via conformal factors}

Let us now derive a formula for the velocity of a star moving on a periodical trajectory in a given galaxy. For simplicity (and in a good agreement
with astronomical observations \cite{Binney}) we will assume the orbit to be circular. In this case the centripetal acceleration and the velocity are related by
\begin{equation}\label{v1}
a=-\frac{v^2}{r}
\end{equation}
On the other hand the Einstein Equivalence Principle remains valid for a theory of gravity that is conformal related with standard GR. This implies
that a test particle (a star in our considerations) will satisfy the geodesic equation
\begin{equation}\label{geod1}
\frac{d^2 x^\mu}{d s^2} + \Gamma^\mu_{\nu\sigma}\frac{d x^\nu}{d s}\frac{d x^\sigma}{d s}=0.
\end{equation}
Stars can move around the galactic center at very high velocities. However, compared with the speed of light, the velocities are still much smaller such that
the condition $v<<c$ is always satisfied. Using the coordinate parametrization $(x^0,x^1,x^2,x^3)=(ct,r,\theta,\varphi)$ it follows immediately that
if $v<<c$ then also $v^i=(dr/dt,rd\theta/dt,r\sin\theta d\varphi/dt)<<dx^0/dt$. Under these conditions, together with the week field limit of the geodesic
equation (\ref{geod1}) for a static spacetime ($\Gamma^0_{00}=0$), one obtains for the radial component
\begin{equation}\label{geod2}
\frac{d^2 r}{d t^2} =-c^2\Gamma^r_{0 0}.
\end{equation}
Inserting now Eq. (\ref{geod2}) into (\ref{v1}) we simply get
\begin{equation}\label{v2}
v^2(r)=rc^2\Gamma^r_{0 0}.
\end{equation}
We have already discussed the idea of projective structures, that is, the class of connections related to each other by Eq. (\ref{connec}). As already mentioned,
connections belonging to the same projective structure describe the same geodesics but differently parameterized. One needs to choose which metric from the conformal
structure is connected to the geodesic motion. Let us consider the case when
one deals with the Weyl geometry: the connection appearing above is a Levi-Civita connection of the conformal metric $\tilde{g}\equiv h$:
\begin{equation}\label{conf}
h_{\mu\nu}=\Omega\,g_{\mu\nu}
\end{equation}
One has that $\Gamma^r_{0 0}$ entering into Eq. (\ref{v2}) is
\begin{eqnarray}
\Gamma^r_{0 0} = \frac{1}{2}h^{r\sigma}(h_{\sigma 0,0}+h_{0\sigma,0}-h_{00,\sigma})=-\frac{1}{2}h^{rj}\,\partial_j h_{0 0}
\end{eqnarray}
which for a spherical-symmetric metric
\begin{equation}
 ds^2=h_{\mu\nu}dx^\mu dx^\nu=-c^2B(r)dt^2+A(r)dr^2+r^2d\tilde{\Omega}^2
\end{equation}
takes the form
\begin{equation}
\Gamma^r_{0 0} =\frac{1}{2}\frac{B'(r)}{A(r)}
\end{equation}
and needs to be computed for a chosen model of gravity. If we consider any modified Einstein field equations of the form (using the convention for $\kappa=-8\pi Gc^{-4}$ from \cite{Weinberg})
\begin{equation}\label{eina}
 \sigma(G_{\mu\nu}-W_{\mu\nu})=\kappa T_{\mu\nu},
\end{equation}
where $\sigma$ represents a coupling to the gravity (for example a scalar field), one can write \cite{an}
\begin{equation}\label{gam}
\Gamma^r_{0 0} =\frac{B(r)}{2A(r)}\left(\frac{A(r)-1}{r}-A(r)rW^r_r\right).
\end{equation}
We assume that the functions $\tilde B(r)=B(r)/\Omega$ and $\tilde A(r)=A(r)/\Omega$ will take the Schwarzschild form in the weak field limit, that is, when we
consider distances much smaller than the core size of a galaxy. In that case, the term $g^{(1)}_{00}\equiv B^{(1)}$ in the approximation
\begin{equation}
h_{\mu\nu} = \Omega(\eta_{\mu\nu}+ g^{(1)}_{\mu\nu} )
\end{equation}
represents the first order correction
coming from the weak field limit of GR (see for example \cite{Weinberg})
\begin{equation}
g^{(1)}_{0 0}=-\frac{2\phi_N}{c^2} = \frac{2GM}{c^2 r}.
\end{equation}
In the next section we will derive the exact form of (\ref{gam}) for a particular model of gravity which admits EPS interpretation.

\subsection{An example: Starobinsky model}\label{secstarob}

In principle there exists an entire class of gravity theories \cite{EPS,Fatibene} that are conformal related with Einstein general
relativity via Eq. (\ref{conf}). Our aim is to explain the observed galaxy rotation curves using Eqs. (\ref{gam}) and (\ref{v2}) without assuming the existence
of Dark Matter.

In what follows we would like to propose a model that fits well the astronomical observed data on galaxy rotation
curves. Our analysis is performed on a subset of galaxies obtained from THINGS: The HI Nearby Galaxy Survey catalogue \cite{Walter,deBlok}, which is a high spectral
and spatial resolution survey of HI emission lines from 34 nearby galaxies.

Any new model must take into account and reproduce the observed flatness of galaxy rotation curves. At short distances (at least the size of the solar system)
the velocity should have as a limit the Newtonian result $v^2(r)=GM/r$. This imposes some constrains on the functions $A(r)$ and $B(r)$.

Before we move further to the Starobinsky model, let us briefly recall the Palatini formalism. The action is
\begin{equation}\label{action}
 S=\frac{1}{2\kappa}\int\mathrm{d}^4x\sqrt{-g}f(\hat{R})+ S_m,
\end{equation}
where $f(\hat{R})$ is a function of a Ricci scalar $\hat{R}=g^{\mu\nu}\hat{R}_{\mu\nu}(\hat{\Gamma})$, while $S_m$ is a matter action
independent of the connection. The Ricci scalar is constructed by
the metric-independent torsion-free connection $\hat{\Gamma}$. Varying the action with respect to the metric gives
\begin{equation}\label{var_met}
 f'(\hat{R})\hat{R}_{(\mu\nu)}-\frac{1}{2}f(\hat{R})g_{\mu\nu}=\kappa T_{\mu\nu},
\end{equation}
were the prime means the differentiation with respect to $\hat{R}$ and as usually $T_{\mu\nu}$ is the standard (symmetric) energy- momentum tensor given by the variation of the matter action $S_m$ with respect to $g_{\mu\nu}$. The $g$-trace
of (\ref{var_met}) arises as the structural equation of the spacetime controlling (\ref{var_met})
\begin{equation}\label{struc}
  f'(\hat{R})\hat{R}-2f(\hat{R})=\kappa T.
\end{equation}
Assuming that we are able to solve (\ref{struc}) as $\hat{R}(T)$ we get that $f(\hat{R})$ is a function
of the energy-momentum tensor trace $T$, that is, $T=g^{\mu\nu}T_{\mu\nu}$.
The variation of (\ref{action}) with respect to the connection is
\begin{equation}
 \hat{\nabla}_\lambda(\sqrt{-g}f'(\hat{R})g^{\mu\nu})=0.
\end{equation}
From the above it immediately follows that the connection $\hat{\nabla}$ is the Levi-Civita connection for the conformally related
metric $f'(\hat{R})g_{\mu\nu}.$ For more detailed discussion we suggest to see for
example \cite{Allemandi:2005qs, Allemandi:2004ca, Allemandi:2004wn}.

The matter part of the modified Einstein equations will be considered as the perfect fluid energy-momentum tensor with the trace
\begin{equation}\label{trace}
 T=3p-\rho.
\end{equation}
In the following work we will consider the pressureless case, that is, for dust $p=0$. Moreover, we will assume only the radial dependence of the energy density, that is,
$\rho=\rho(r)$.

Let us notice \cite{dicke} that Eq. (\ref{var_met}) can be transformed into so-called Einstein frame
\begin{equation}
 \tilde{G}_{\mu\nu}=\kappa \tilde{T}_{\mu\nu}-\frac{1}{2}h_{\mu\nu}\tilde{U}
\end{equation}
where the Einstein's tensor is constructed with the conformal metric $h_{\mu\nu}$ while $\tilde{T}_{\mu\nu}=T_{\mu\nu}/f'(\hat{R})$
and the effective potential $\tilde{U}=f'(\hat{R})^{-2}(\hat{R}f'(\hat{R})-f(\hat{R}))$. Thus, we see that
\begin{equation}
 f'(\hat{R})\Big(\tilde{G}_{\mu\nu}+\frac{1}{2}h_{\mu\nu}\tilde{U}\Big)=\kappa T_{\mu\nu}
\end{equation}
can be treated in similar manner as Eq. (\ref{eina}) with $W_{\mu\nu}=-\frac{1}{2}h_{\mu\nu}\widetilde{U}$.

Now, we are equipped with all the tools needed in order to examine the model due to our assumption on the dynamics. As was already mentioned,
the easiest modification of the Einstein-Hilbert action is adding the Starobinsky quadratic term \cite{starob}:
\begin{equation}
f(\hat{R})=\hat{R}+\gamma \hat{R}^2,
\end{equation}
where $\gamma$ is a very small parameter having a significance in the case of a strong gravitational field. From the cosmological consideration of the model in Ref.
\cite{bor3}, with cosmological constant added additionally to the matter part, one gets that $\gamma$ is of order $10^{-11}$.
Since in the considered case the conformal factor is simply $\Omega=f'(\hat{R})=1+2\gamma \hat{R}$ and that the structural equation (\ref{struc}) for the dust
matter gives rise to
\begin{equation}
 \hat{R}=-\kappa T= c^2\kappa\rho,
\end{equation}
we get particularly for the Starobinsky model that
\begin{equation}
 \Omega=1+2\kappa c^2\gamma\rho.
\end{equation}
Using the results from \cite{an} one finds that
\begin{equation}\label{sol}
 A(r)=\left( 1-\frac{2GM(r)}{c^2r} \right)^{-1}
\end{equation}
where the modified mass distribution has the following form
\begin{equation}\label{mod_mass}
 M(r)=\int^r_0\frac{4\pi\tilde r^2\rho(\tilde{r})}{1+2c^2\kappa\gamma\rho(\tilde r)}\left(1+
 \frac{c^2\kappa\gamma\rho(\tilde r)}{2(1+2c^2\kappa\gamma\rho(\tilde r))} \right)d\tilde r.
\end{equation}
Let us now assume a simple galaxy model obtained from the following matter distribution
\begin{equation}\label{massnew}
\tilde M(r)=M_0\left( \sqrt{\frac{R_0}{r_c}}\frac{r}{r+r_c} \right)^{3\beta}
\end{equation}
where $r_c$ can be interpreted as the "core radius", $M_0$ is the total mass of the galaxy and $R_0$ is the scale length of the galaxy to be matched with
the observed one. Like in Ref. \cite{Moffat2}, the values of the parameter $\beta$ is $1$ for high surface brightness galaxies (HSB) and $\beta=2$
in the case of low surface brightness galaxies (LSB).
This matter distribution is a slightly modified version of the model used in Ref. \cite{Moffat2} in order to be more close to the actual mass
profile as inferred from the observed photometric profile. Although Eq. (\ref{massnew}) is still a very rough approximation it
is good enough for the purposes of this paper.

It can be noticed that the mass distributions (\ref{mod_mass}) and (\ref{massnew}) can be identified for a suitable choice of energy density $\rho(r)$ in (\ref{mod_mass}); more exactly,
comparing the derivatives $\tilde M'(r)=M'(r)$ one gets algebraic equation for $\rho(r)$.

Moreover, with the solution (\ref{sol}) we are able to find that formula (\ref{gam}) is now
\begin{equation}
 \Gamma^r_{00}=\frac{B(r)}{2}\left( \frac{2GM(r)}{c^2r^2}+\frac{\kappa^2\gamma c^4r\rho^2}{2(1+2\kappa\gamma c^2\rho)^2} \right)
\end{equation}
and therefore we write the quadratic velocity as
\begin{equation}
 v^2=\frac{GB(r)M(r)}{r}\left( 1-\frac{2\pi\kappa\gamma c^2r^3\rho^2}{M(r)(1+2\kappa\gamma c^2\rho)^2} \right).
\end{equation}
Since the exact form of $B(r)$ is very complex, let us take the approximated value, that is,
\begin{equation}
 B(r)=\Omega(1+ g^{(1)}_{00})\approx 1+ g^{(1)}_{00}=1+\frac{2GM(r)}{c^2r}
\end{equation}
It follows immediately that the circular velocity of a star around the galactic center can be approximated by
\begin{equation}\label{vstarob}
  v^2\approx \frac{GM(r)}{r}\left( 1+\frac{2GM(r)}{c^2r}-\frac{2\pi\kappa\gamma c^2r^3\rho^2}{M(r)(1+2\kappa\gamma c^2\rho)^2} \right).
\end{equation}
The above formula (\ref{vstarob}) is the main result of this paper, that is, the circular velocity obtained from the Starobinsky lagrangian in Palatini formalism. It is now used to obtain plots for a sample of 6 HSB galaxies after determining
the parameters $M_0$ and $r_c$ using the {\verb NonlinearModelFit } function in {\verb WolframMathematica }.
They are presented in Figure \ref{fig.1}.

\begin{table}
\renewcommand{\arraystretch}{1.3}
\caption{Best fit results according to Eq. (\ref{vstarob}) using the parametric mass
distribution (\ref{massnew}). These numerical values correspond to rotation curves presented in Fig. \ref{fig.1}. Col. (1) galaxy name; Col. (2) total gas mass, in units of $10^{10}\,M_{\odot}$, given by $M_{gas}=4/3M_{HI}$, with the $M_{HI}$ data taken from \cite{Walter}; Col. (3) measured scale length of the galaxy in kpc; Col. (4) galaxy luminosity in the B-band, in units of $10^{10}\,L_{\odot}$, calculated from \cite{Walter}; Col. (5) presents the best-fit results for the predicted total mass of the galaxy $M_0$ (in $10^{10}\,M_{\odot}$ units); col. (6) gives the predicted core radius $r_c$ in kpc; Col. (7) reduced $\chi^2_r$; and Col. (8) the stelar mass-to-light ratio in units of $M_{\odot}/L_{\odot}$ . Note: all 6 galaxies are type HSB. }
\label{tab1}
\begin{tabular*}{\textwidth}{cccccccc}
\cline{1-8}
 Galaxy & $M_{gas}$ & $R_0$ & $L_B$ & $M_0$ & $r_c$ & $\chi^2_r$ & $M/L$ \\
 (1) & (2) & (3) & (4) & (5) & (6) & (7) & (8) \\ \cline{1-8}
 NGC 3031 & 0.48 & 2.6 & 3.049 & 14.86 & 2.10 & 4.88 & 4.71 \\
 NGC 3521 & 1.07 & 3.3 & 3.698 & 38.45 & 3.69 & 1.84 & 10.10 \\
 NGC 3627 & 0.11 & 3.1 & 3.076 & 8.68  & 2.25 & 0.45 & 2.78 \\
 NGC 4736 & 0.05 & 2.1 & 1.294 & 0.53  & 0.59 & 2.41 & 0.37 \\
 NGC 6946 & 0.55 & 2.9 & 2.729 & 78.19 & 5.09 & 2.18 & 28.44 \\
 NGC 7793 & 0.12 & 1.7 & 0.511 & 18.24 & 3.36 & 4.82 & 35.45 \\
 \cline{1-8}
\end{tabular*}
\end{table}

\begin{figure}[h!t]
\centering
\includegraphics[scale=0.4]{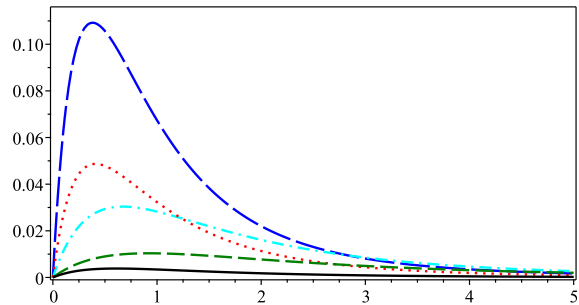}
\caption{The difference $v-v_{newt}$ (vertical axis in Km/s) as a function of distance (horizontal axis in kpc), where $v$ is given
by eq. (\ref{vstarob}) and $v_{newt}$ is the Newtonian velocity $GM(r)/r$. The galaxies from top to bottom are as follows: NGC3031, NGC3627,
NGC3521, NGC6946 and NGC7793. The galaxy NGC4736 is not shown, but for it always $v-v_{newt}<0.7$ }
\label{fig.3}
\end{figure}

\begin{figure*}[h!t]
\centering
\includegraphics[scale=0.293]{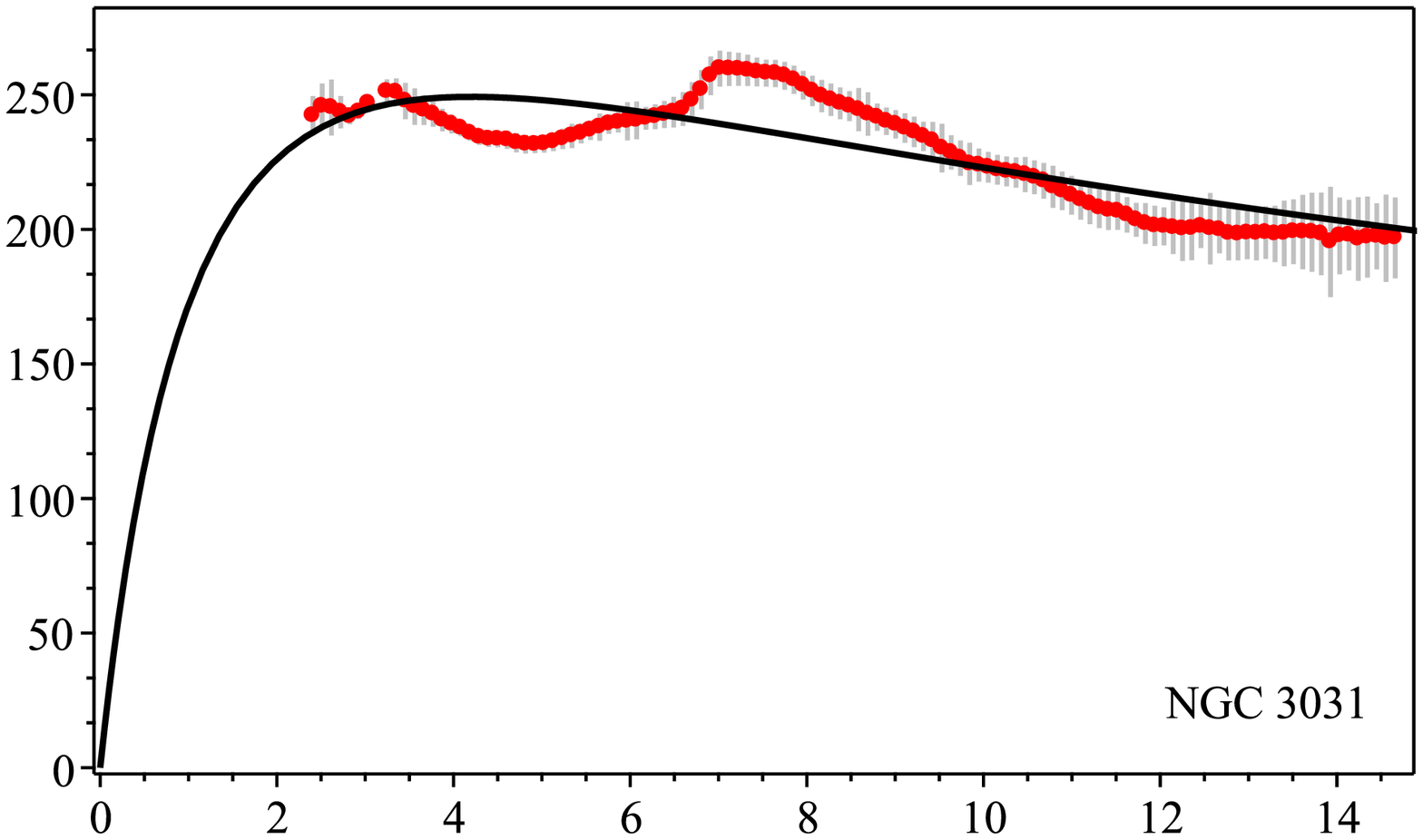}
\includegraphics[scale=0.293]{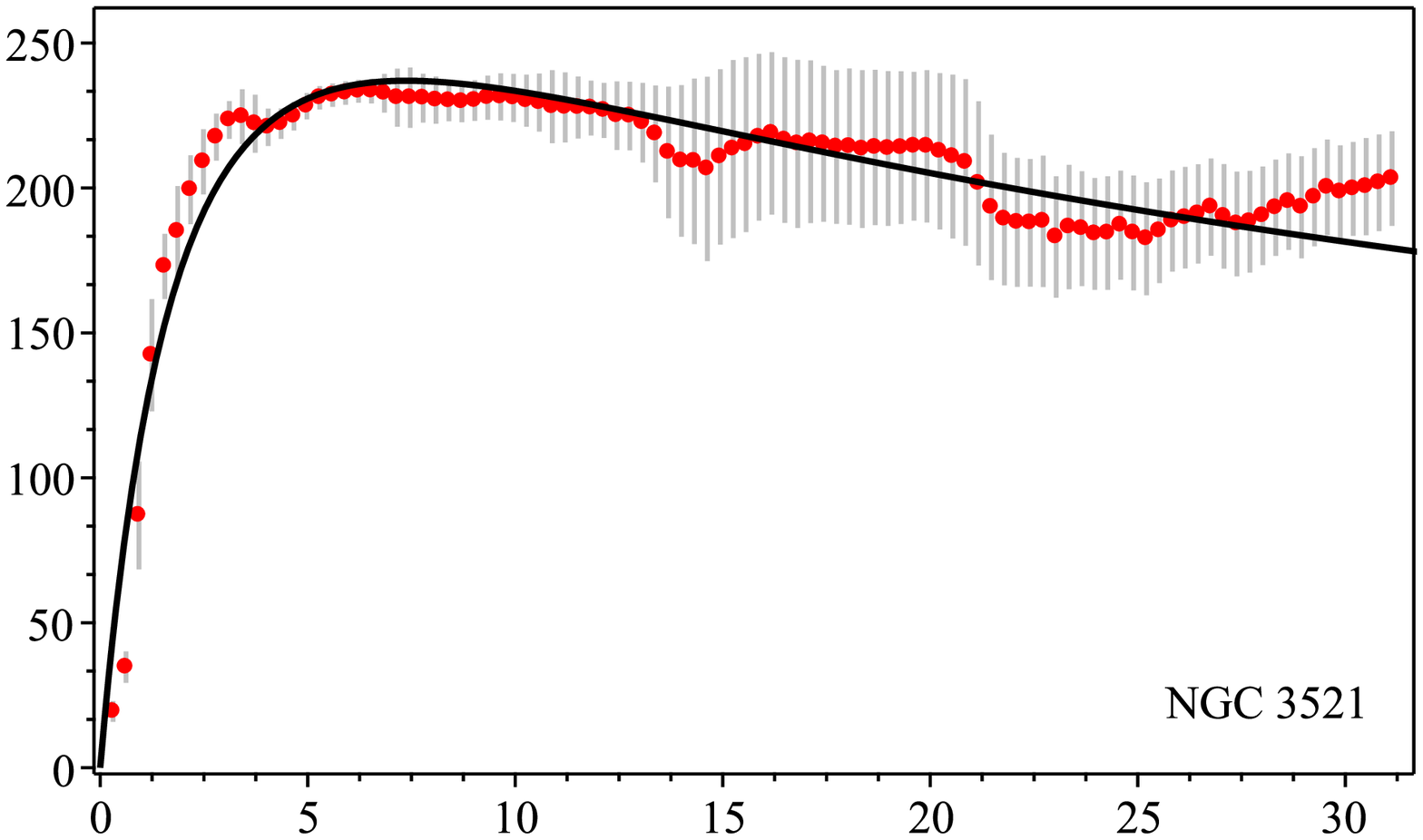}
\includegraphics[scale=0.293]{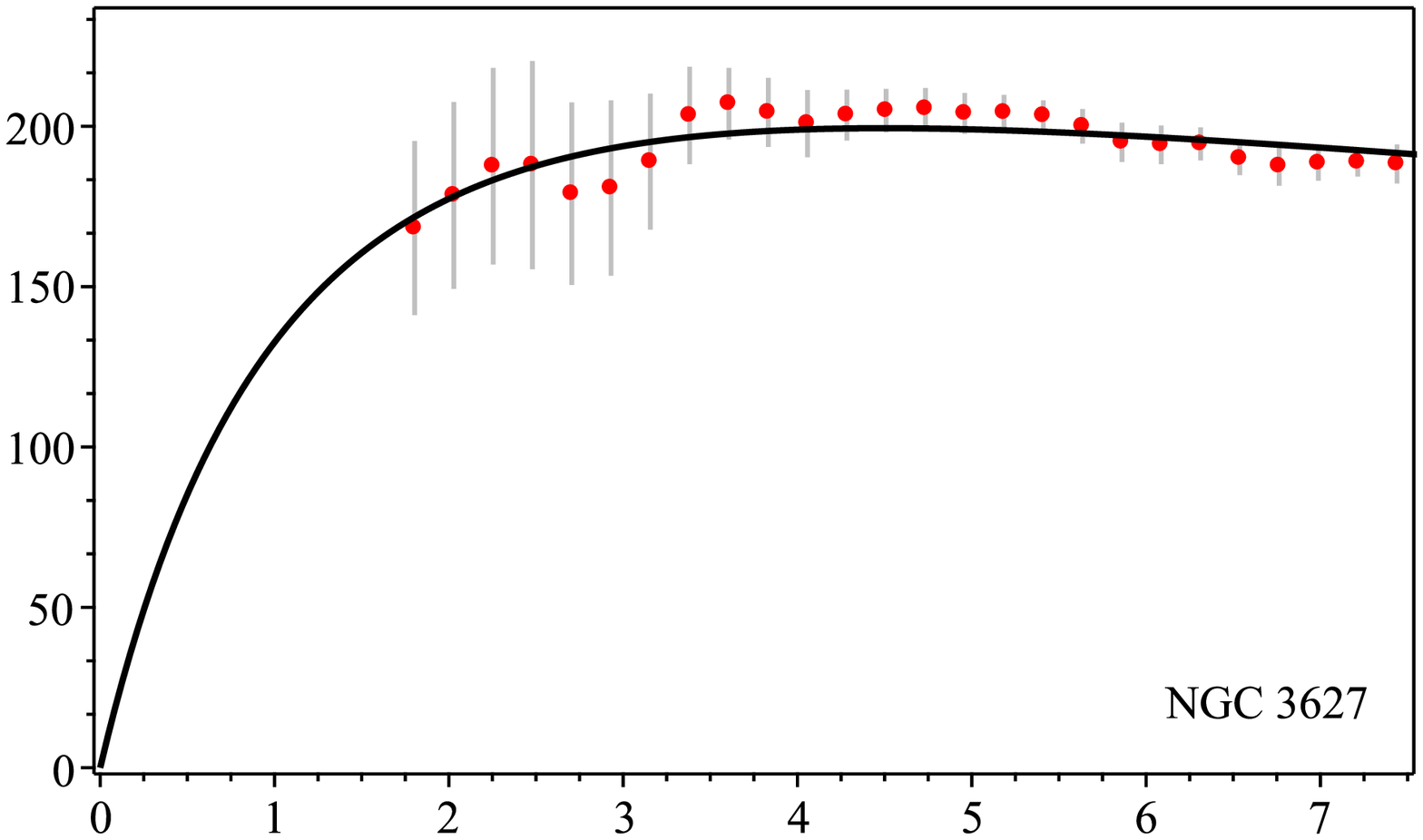}
\includegraphics[scale=0.293]{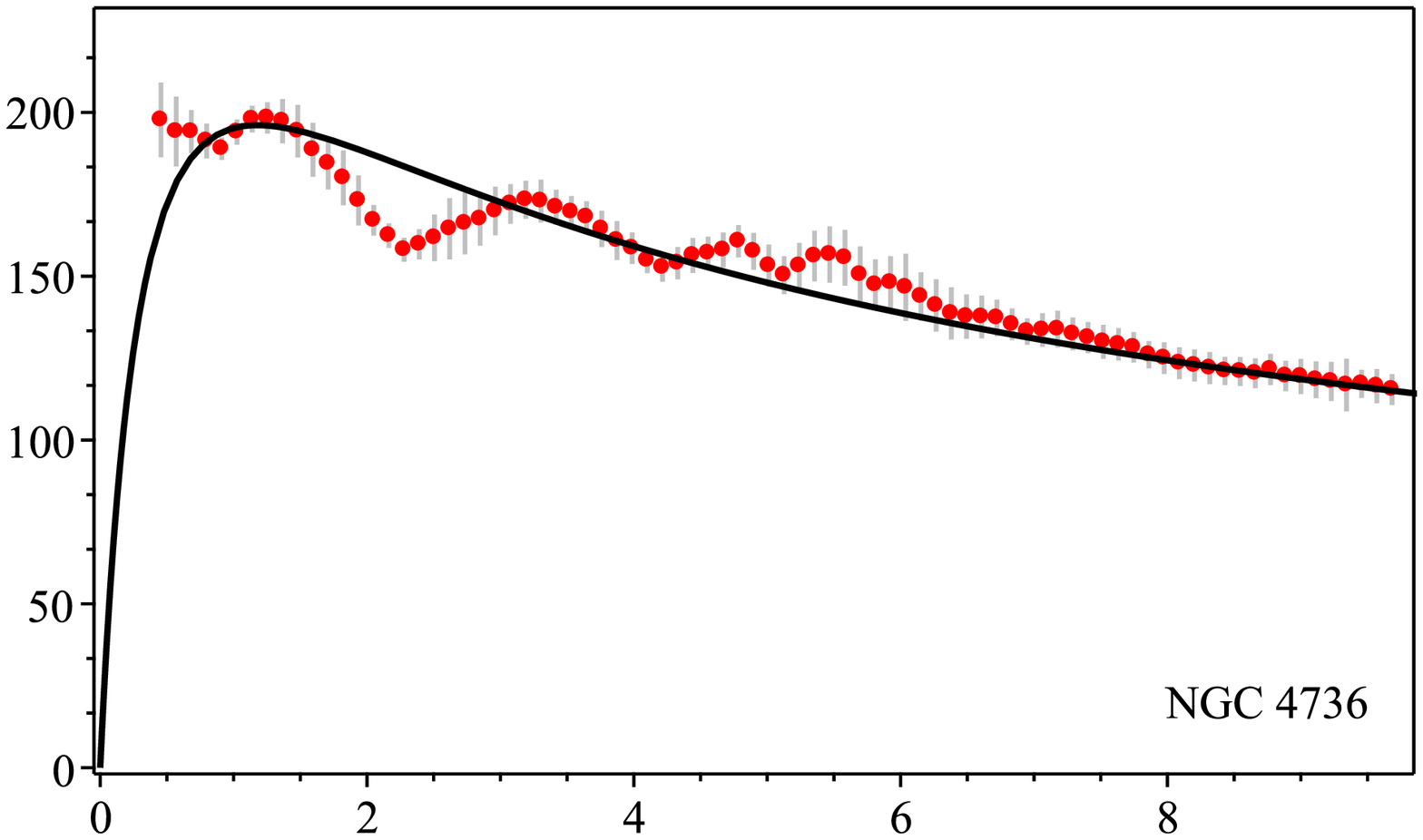}
\includegraphics[scale=0.293]{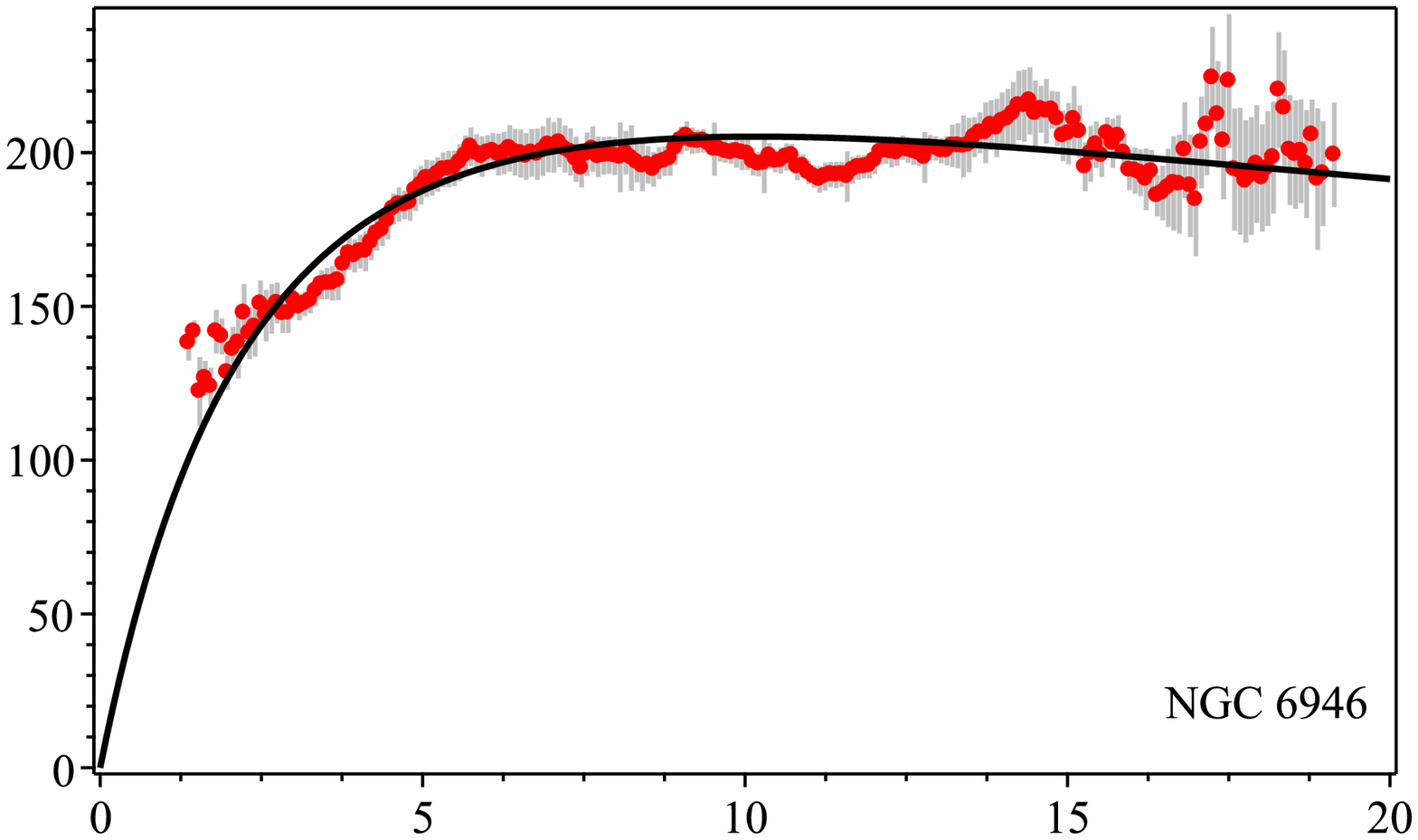}
\includegraphics[scale=0.293]{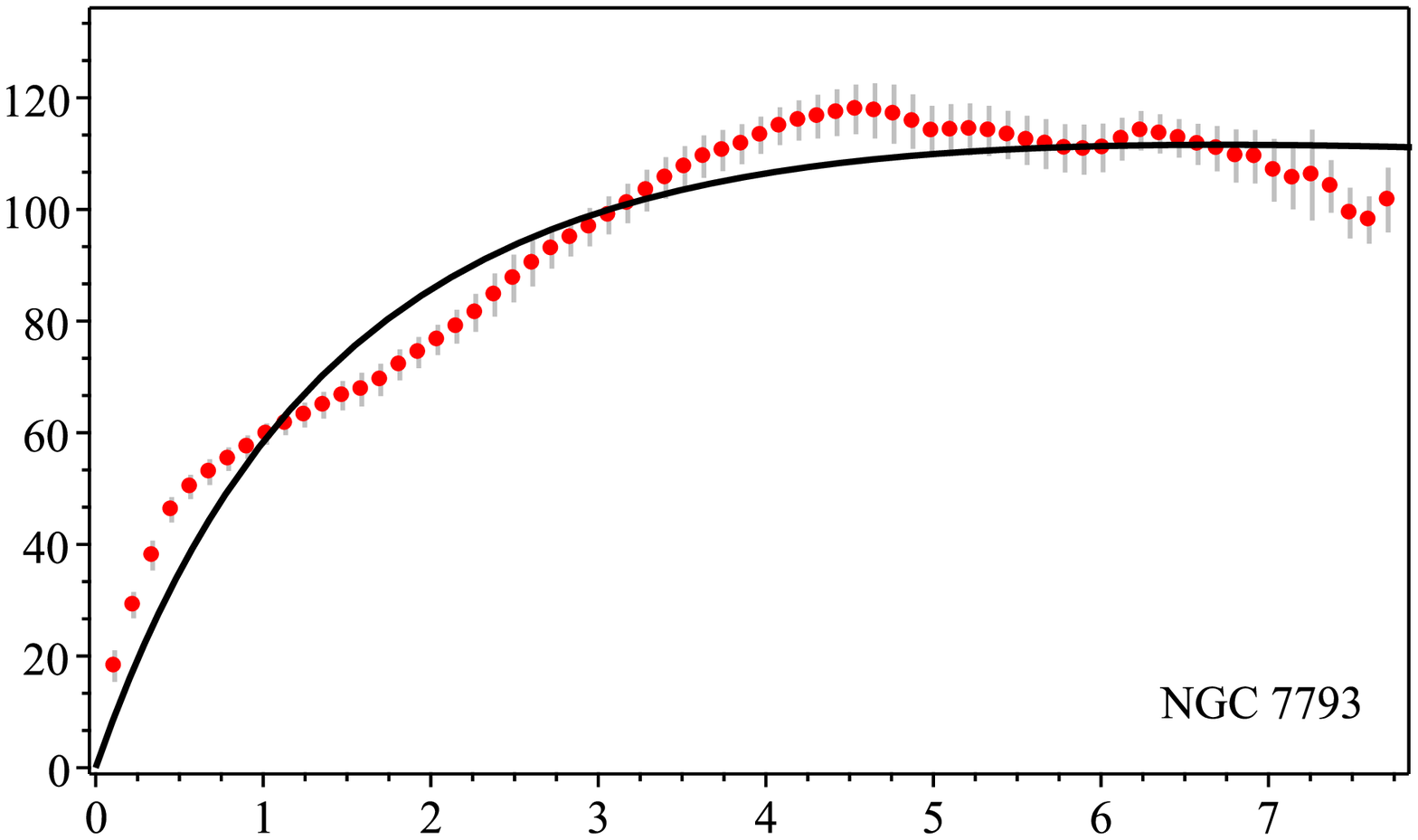}
\caption{(color online) Rotational velocities (in km/sec) as a function of distance (in kpc). The black curve represents the best
parametric fit of galaxy rotation curves using eq. (\ref{vstarob}) for the sub-sample of 6-THINGS galaxies. The values of the best-fit
parameters can be found in Table \ref{tab1}.}
\label{fig.1}
\end{figure*}

The corresponding values for $M_0$ and $r_c$ are presented in Table \ref{tab1}. We observe a very good agreement between the data
points and the fitted continuous black curve. Unfortunately, after plotting the Newtonian curves using the same
values of $M_0$ and $r_c$ from Table \ref{tab1} we have found that there is almost no difference (see Figure \ref{fig.3})
between the Newtonian curve and the one derived using Eq. (\ref{vstarob}). Furthermore, the mass-to-light $M/L$ values
inferred from the fit in Table \ref{tab1} are also too large compared to what is expected based on stellar population
synthesis models \cite{McGaugh}.

\section{Conclusions}
In this paper we have considered the possible explanation of observed galaxy rotation curves by the assumption that the metric and the
connection are independent objects in the spirit of EPS formalism. We studied the case when the connection is a Levi-Civita connection of a metric conformally
related to the metric which is responsible for the measurement of distances and angles. Due to that interpretation masses moving in a gravitational field should
follow geodesics appointed by the connection providing different equations of motion. It turns out that the rotational velocity formula obtained under this
formalism differs from the Newtonian one by the presence of extra terms coming indirectly from the conformal factor of the metrics. This term is treated as a deviation from the
Newtonian limit of General Relativity.

In Section \ref{secstarob} we used Palatini gravity, which is a representation of the EPS formalism, and as a working example
we took the Starobinsky Lagrangian $f(\hat{R})=\hat{R}+\gamma \hat{R}^2$ in order to derive a rotational velocity formula given by
the expression in Eq. (\ref{vstarob}) for a star moving in a circular
trajectory around the galactic center. Our results are presented in Table
 \ref{tab1} together with Figures \ref{fig.3} and \ref{fig.1}. Although the galaxy masses resulted from the fitting of the data sub-sample  proved to be
too high, giving thus rise to unsatisfactory values for the mass-to-light ratio, non the less we have showed that the approach of obtaining galaxy rotation curves via conformal factors can be valid. It should be also noticed that we have used a very simple matter distribution (\ref{massnew}) in
order to be able to obtain an expression for the energy density $\rho(r)$. More complex distributions could possibly give corrections which
would provide different mass-to-light $M/L$ values. This task is one of our future works.

With those, we would like to briefly conclude by saying that the approach of obtaining galaxy rotation curves using two conformally related metrics can be valid and deserves further investigations. Furthermore, by trying other Lagrangians and other gravity models in the future works   it is definitely possible to improve the findings reported here.

\section*{Acknowledgements}

This work made use of THINGS, "The HI Nearby Galaxy Survey" (Walter et al. 2008).
We would like to thank Professors Fabian Walter and Erwin de Blok for helping us in obtaining the RC data from the THINGS catalogue.

AW is partially supported by the grant of the National Science Center (NCN) DEC- 2014/15/B/ST2/00089.
CS was partially supported by a grant of the Ministry of National Education and Scientific Research, RDI Programme for Space Technology and Advanced Research - STAR, project number 181/20.07.2017.
We appreciate S. Odintsov , D.C. Rodrigues  and S. Vagnozzi for drawing our attention to their papers.
This paper is based upon work from COST action CA15117 (CANTATA), supported by COST (European Cooperation in Science and Technology) \\

\end{document}